# Topological Magnetic Hysteresis in Single Crystals of CeAgSb$_2$ Ferromagnet

Ruslan Prozorov, Sergey L. Bud'ko and Paul C. Canfield

*Ames Laboratory and Department of Physics and Astronomy, Iowa State University, Ames, Iowa 50011*

Date: 11 June 2021

**Closed-topology magnetic domains are usually observed in thin films and in an applied magnetic field. Here we report the observation of rectangular cross-section tubular ferromagnetic domains in single thick crystals of CeAgSb$_2$ in zero applied field. Relatively low exchange energy, small net magnetic moment, and anisotropic in-plane crystal electric fields lower the domain wall energy and allow for the formation of the closed-topology patterns. Upon cycling the magnetic field, the domain structure irreversibly transforms into a dendritic open-topology pattern. This transition between closed and open topologies results in a "topological magnetic hysteresis" – the actual hysteresis in magnetization, not due to the imperfections and pinning, but due to the difference in the pattern morphology. Similar physics was suggested before in pure type-I superconductors and is believed to be a generic feature of other nonlinear multi-phase systems in the clean limit.**

The interpretation of the experimental data in ferromagnetic samples is often difficult because the response is determined by both the microscopic physics of the exchange interactions and the macroscopic physics of the dipolar magnetic fields surrounding always-finite samples. In fact, the low-field behavior of magnetization is determined mainly by the latter [1]. It is well-known that in macroscopic ferromagnetic systems, the minimization of the dipolar magnetic field energy leads to the formation of ferromagnetic domains where the local magnetization points in different directions, thus reducing the amplitude of the outside field beyond some distance from the surface. The price to pay for this reduction is the domain wall energy, and the competition between these two contributions to the total energy determines the pattern morphology and the size of the domains. When an external magnetic field changes, domain walls move, leading to the expansion of the domains where local magnetization points in the applied field direction. Similar phenomena occur in ferroelectrics [2] and superconductors [3,4]. In all cases, the macroscopic behavior is described by the effect of demagnetization - an additional magnetic field proportional to the sample magnetic moment [2,5], strictly applicable to ellipsoids but extendable to the arbitrary shape in a straightforward manner [5,6].

Considering the symmetry with respect to magnetization reversal and the necessity to move the domain wall in response to the magnetic field, open topology patterns (laminar, labyrinth, or stripes) are usually observed in bulk ferromagnets where domain wall energy is large [1]. However, in thin ferromagnetic films with perpendicular to the film plane easy axis, closed topologies (bubbles or tubes) are sometimes more favorable. Such patterns (e.g., CMD = cylindrical magnetic domains) have attracted significant attention due to their use in magneto-optical memory devices. In general, the difference between the energies of tubular and stripe patterns is small and depends on the details of the material parameters, such as magnetic anisotropy, exchange energy, and geometry of the sample [1]. The stability of these bubbles or tubes requires not only thin-film geometry but also an applied biasing magnetic field that lifts the above-mentioned orientational degeneracy. In this case, a dominant domain with the magnetization along the applied bias magnetic field surrounds the other type of domain (with magnetization oriented opposite to the applied field) that may exist in the form of separate flux tubes. At zero applied field, such topological imbalance was believed to be possible only as a metastable state, which can be quenched by annealing above the Curie temperature with subsequent cooling in zero field.

In this contribution, we describe a fascinating example of a closed topology (tubular) pattern in a zero magnetic field that still preserves the topological symmetry on the scale of the entire sample. Remarkably, this is observed not in a thin film but a single thick crystal of CeAgSb$_2$.

In general, the understanding of the domain patterns and their evolution with changing the external parameters is essential for many reasons. Not only it involves the microscopic physics of the material, but it is also directly linked to the coercivity and magnetic losses in the varying magnetic field. The measured magnetization is a sum of the projections of all local magnetic moments onto the measurement axis. Therefore, different M(H) magnetization loops will be recorded for different patterns. The magnetic hysteresis is more often than not observed in ferromagnetic samplers, and it is related to the magnetic domains and their motion. Usually, it is attributed to the imperfections of the crystal lattice and various impurities (collectively known as the "pinning centers") that lead to the position-dependent energy landscape that leads to the pinning of the domain walls and prevents them from moving. However, a different kind of so-called "topological hysteresis" has been demonstrated in clean type-I superconductors, where it results from the differences in the topology of the intermediate state patterns for increasing and decreasing magnetic fields [7-9]. Is it possible to have a similar phenomenon in a ferromagnetic system? As we show here, the answer is a resounding yes.

Overall, this subject is related to a more general discussion of pattern formation in highly nonlinear systems. There are many examples in chemistry, biology, astrophysics, mathematical complexity, and virtually all fields of science [10-13]. Ferromagnetic and superconducting materials are particularly attractive, because electronic and magnetic patterns can be easily manipulated and tuned by varying temperature, magnetic field, and specimen geometry. They have no inertia and are easy to reset, compared, for example, to real-life diffusive mass transport, irreversible chemical reactions, and froth coarsening. Magnetic systems can be tuned as close to the minimum energy state as experimentally possible.

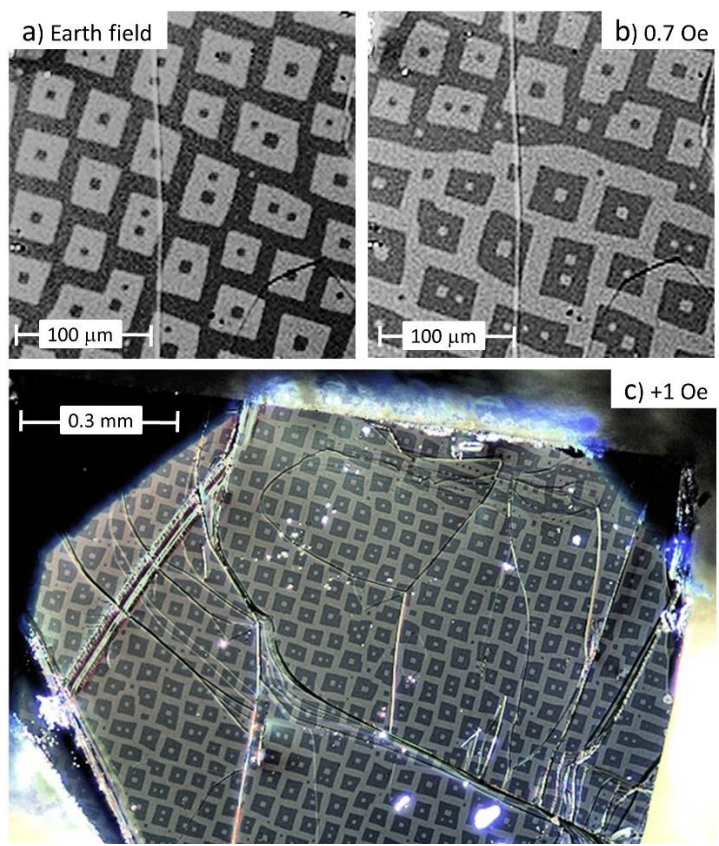

Figure 1. All images are taken at T=5 K. (a) in a nominal Earth magnetic field; (b) cooling in 0.7 Oe field to compensate the Earth magnetic field; (c) the entire sample is cooled in 1 Oe field.

CeAgSb$_2$ is a ferromagnetic moderately heavy-fermion compound with Curie temperature, $T_c$=9.8 K, and a small net magnetic moment of about 0.4$\mu_B$ on Ce$^{3+}$ sites [14-18]. Spontaneous magnetization points in the c-axis direction, which was understood within the ferromagnetic ground state in anisotropic exchange scheme [17,18]. Remarkably, there is competing, *ab*-plane easy axes, dictated by the crystal electric field (CEF) at least in the paramagnetic state [17]. This may have profound implications on the domain pattern. Specifically, relatively small exchange energy (small $T_c$), small net magnetic moment, and perpendicular to magnetization direction CEF easy axis all point toward relatively small domain wall energy, which favors the tubular domain structure in a system with uniaxial magnetic anisotropy.

The extremely soft nature of the magnetic domains in CeAgSb$_2$ became apparent when we realized that even the Earth's magnetic field was sufficient to alter the domain pattern! In the first set of experiments, we observed the structure shown in Fig.1(a). However, there is an apparent problem with this picture - such domain pattern breaks the topological symmetry in the sense that one type of domains is not topologically equivalent to the opposite-sign domains, and this is the case on the scale of the entire sample, as shown in Fig.1(c). Clearly, there is a reason to break the expected topological invariance and, as is often the case with a very soft magnetic response, the problem is the Earth's magnetic field. Yet, this is quite extraordinary that the domain structure can be altered by a minuscule 0.5 Oe magnetic field! With $T_c = 9.8\,\text{K}$, the characteristic field at which magnetostatic energy equates to thermal energy is $B = k_B T/0.4\mu_B \approx 36.5\,\text{T}$, almost a million times greater than the magnetic field required to move the domains. Indeed, by applying a precisely controlled magnetic field, we could recover the topological invariance with respect to the direction of magnetization in the neighboring domains. However, as shown in Fig.1(b), this symmetry is multiscale – two types of domains are observed. Each large domain contains closed-topology domains of the opposite orientation, and, together, the overall magnetic structure is symmetric with respect to the field reversal. Cooling in a slightly larger magnetic field, + 1 Oe, turns the whole picture, Fig.1(c), into a negative image of Fig.1(a). This highly unusual compensation at the scale of the entire sample signifies the contribution of the far regions around the sample to the overall magnetostatic energy.

Clearly, the observed domain structure is more complex than simple circular domains induced in garnets and other films. For example, in Fig.1(a), the dark continuous domain contains the opposite sign, bright domain, as rectangular cross-section regions with rectangular and round dots in the center. Unfortunately, we do not have a way to look into the bulk and determine whether these patterns are continuous throughout the thickness or only appear at the surface due to so-called domain branching [1]. A study similar to what was done in Nd$_2$Fe$_{14}$B with bulk scattering spectroscopy needs to be performed to answer this question [19]. In our opinion, it is most likely that the rectangular regions are continuous flux tubes with very low domain wall energy so that they follow the in-plane crystal field anisotropy in a tetragonal crystal. Their evolution is reflected in the M(H) loops described below, causing substantial changes in the total magnetization. If they were just surface features, the total magnetization would not be affected nearly that much. Yet, it is possible that the central dots in the rectangular may be the surface pockets that provide additional near-field magnetostatic compensation.

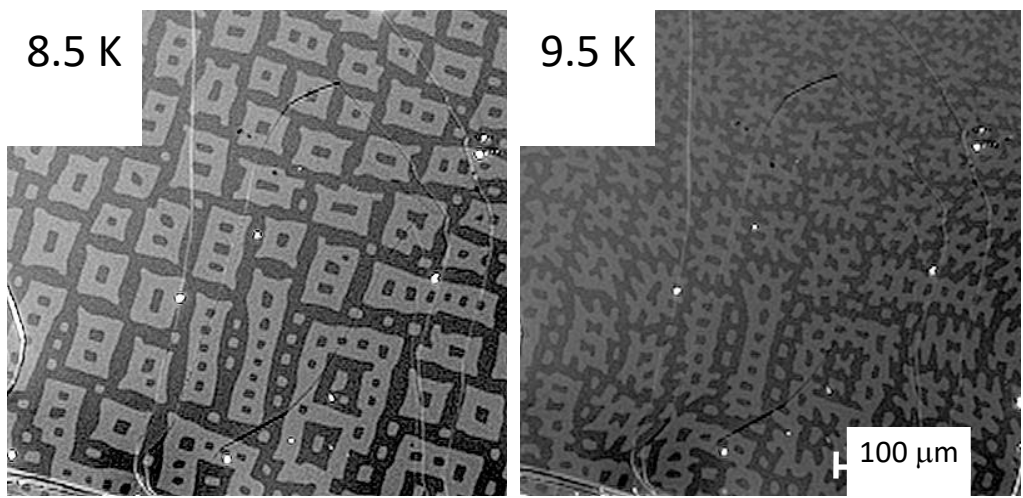

Figure 2. Transformation of the domain structure upon warming approaching $T_c$.

To the best of our knowledge, this is the first observation of such a domain pattern when bulk ferromagnet splits into super-domains, each containing the rectangular tubes of one spin orientation embedded in a matrix of the opposite spin. When the external magnetic field is strictly zero, the total magnetization is zero, and topological symmetry is preserved on the scale of the entire sample. However, on a scale of each super-domain, the tubular structure is favored probably because domain wall energy is small (compared to typical ferromagnets). This is similar to the behavior of the suprafroth in the superconducting lead, which is a moderate type-I superconductor with very low energy of the superconductor/normal phase interface [7-9,11].

The observation of the magnetic domains is usually performed well below the Curie temperature without monitoring their transformation across the transition, which is often too high.

Here we show that it is imperative to be able to zero-field cool the sample, because otherwise the equilibrium domain structure may never be attained. In CeAgSb$_2$, $T_c$=9.8 K, and we could observe these transformations. Figure 2 shows the change in the magnetic domain structure upon warming up after cooling in Earth magnetic field. Comparison of Fig.1(a) and Fig.2(a) shows that little change occurs between 5 K and 8.5 K. However, the domains become more rounded and branch out closer to Tc. Both effects can be linked to a further decrease of the magnetic anisotropy along the c-axis and, perhaps, the desire of the ordered moment to follow the in-plane CEF. Importantly, when the magnetic field is cycled at temperatures below $T_c$, the domain structure changes and never recovers. It makes us question: how many possible domain patterns were missed in the ferromagnetic materials with $T_c$ too high for such an experiment?

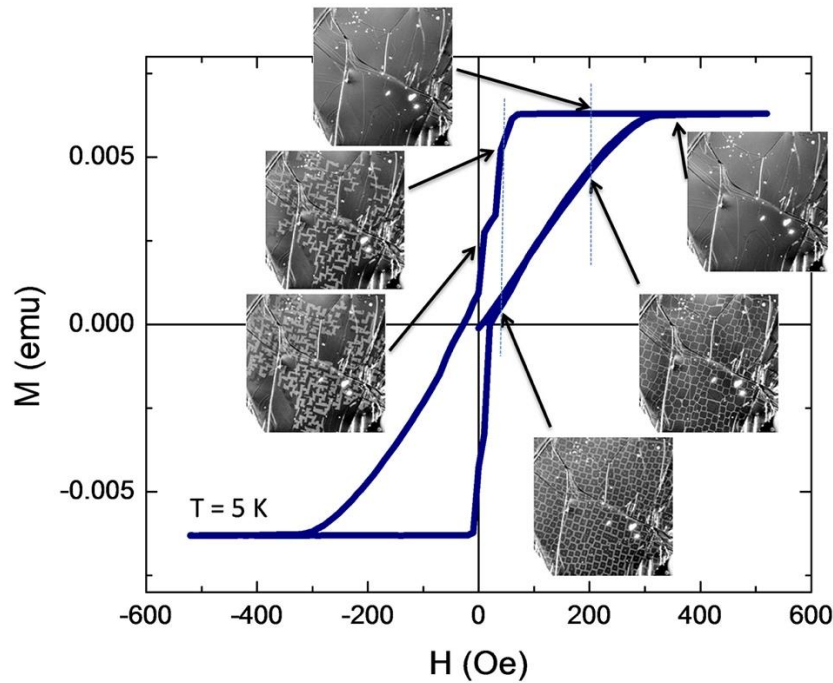

Figure 3. Magnetization loop measured at 5 K illustrated by the images of the domain pattern at the indicated magnetic fields.

Figure 3 shows the evolution of the measured total magnetic moment upon an increase of the applied field after zero-field cooling, reaching beyond the saturation magnetization, reducing back to negative magnetic fields, and finally returning back to larger positive magnetic fields. The corresponding characteristic images of the evolution of the magnetic domains are shown. They start as a quite regular domain pattern described above, evolve into a monodomain fully magnetized state, but return to the dendritic network of domain walls crossing each other at right angles, perhaps following the CEF anisotropy. Obviously, there is a magnetic hysteretic behavior related to the difference in the domain topology, similar to the "topological hysteresis" of type-I superconductors. This is further investigated in Fig.4, which examines the magnetic field dependence of magnetization obtained in different ways.

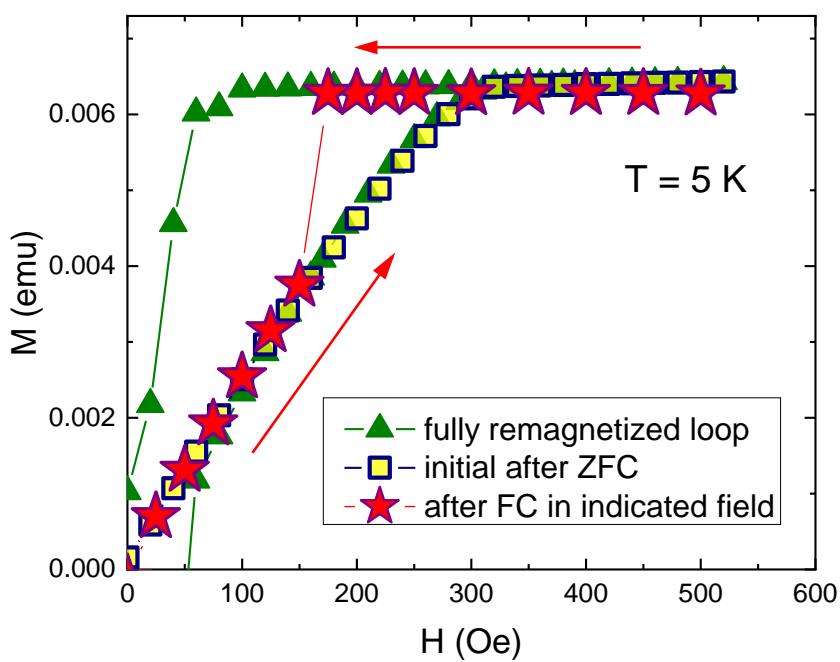

Figure 4. Pristine initial magnetization curve after cooling in zero magnetic field (ZFC) (yellow squares) and a returning branch after full remagnetization at the negative magnetic field are compared to a field-cooled magnetization performed at the indicated magnetic fields (red stars).

Here we arrived at each point on the M(H) loop following three different paths in the T-H phase diagram. The first two are the conventional M(H) hysteresis loops measured at a constant temperature, shown in Fig.4 by triangles and squares. The yellow squares show the "initial" magnetization measured in an increasing magnetic field after cooling in zero magnetic field (ZFC) to T=5 K. This path is accompanied by the gradual transformation of the tubular domain structure shown in Fig.3. The green triangles in Fig.4 show the same branch of the full M(H) loop after remagnetization in large positive and large negative magnetic fields. The third protocol, shown by the red stars, is M(H) measured after cooling in the indicated magnetic field from above Tc to 5 K. Obviously, after the metastable uniform state, an equilibrium tubular structure is triggered below 200 Oe, and the magnetization there matches exactly that obtained after ZFC pristine state. This provides crucial evidence that the tubular structure is the equilibrium state of this ferromagnetic system. The difference between green triangles in the descending branch and red circles is the topological hysteresis in the sense that the macroscopic magnetization hysteresis depends on the topology of the domain pattern, which in turn is determined by the path to a given point in the T-H phase diagram.

In conclusion, we discovered the rectangular tubular pattern of the magnetic domains in thick ferromagnetic crystals of $CeAgSb_2$, which represent the equilibrium (minimum energy) topology. This result proves that such closed-topology patterns are not the property of thin-film geometry, applied magnetic field, or the result of a particular metastable state but may be formed if domain wall energy is unusually small and there are competing anisotropic contributions to the total magnetostatic energy.

### Acknowledgments

This research was supported by the U.S. Department of Energy, Office of Science, Basic Energy Sciences, Materials Science and Engineering Division through the Ames Laboratory. The Ames Laboratory is operated for the U.S. Department of Energy by Iowa State University under Contract No. DE-AC02-07CH11358.

## Materials and Methods

Large single crystals of CeAgSb$_2$ were grown out of excess Sb. High-purity elements were placed in alumina crucibles with an atomic ratio of Ce: 0.045, Ag: 0.091, Sb: 0.864 and sealed in evacuated amorphous silica tubes. The ampules were heated to 1180 °C and then cooled over 100 hours to 670 °C after which the excess Sb was decanted with the help of a centrifuge [20]. Single crystals with masses as large as ~2 grams were grown (see reference [15] for a representative picture). Detailed structural, magnetic, transport, and thermodynamic measurements were reported previously [14,17].

Direct magneto-optical Kerr imaging was used to reveal magnetic domain structure. Specifically, the sample is mounted on a copper cold stage with its natural (as grown) [001] facet perpendicular to the incident linearly polarized light. The copper cold stage is situated inside the optical flow-type liquid helium cryostat. The c-axis (also easy magnetic axis) is parallel to the light propagation direction and is therefore optimal for the magneto-optical polar Kerr effect. Upon reflection, linear polarization direction (parallel to the surface) rotates by the angle proportional to the surface magnetization. Due to the chirality of the problem, opposite magnetic moments lead to the opposite directions of the polarization rotation. When viewed through an analyzer rotated almost perpendicular to the polarizer, a 2D image of the magnetic pattern emerges. In all images, bright regions correspond to UP domains, whereas dark regions correspond to DOWN domains. A more detailed discussion of magneto-optical techniques can be found elsewhere [1,3].

Magnetic measurements were conducted by using a commercial 5 T *Quantum Design* Magnetic Property Measurement System (MPMS).